\documentclass{menu99}    % A4 paper size
\usepackage{epsfig}
\usepackage{tabularx}

\newcolumntype{Y}{>{\small\raggedleft\arraybackslash}X}
\newcommand{\st}{\scriptstyle}

\newcommand{\cl}{\centering}

\begin{document}
    \setlength{\baselineskip}{2.6ex}

\rightline{UNITUE--THEP--14/99}
\rightline{nucl-th/9911020}

\title{Kaon Photoproduction  in a Confining and Covariant Diquark
Model\thanks{Supported by the BMBF (06--TU--888) and by the DFG (We 1254/4-1).
\newline Talk given by R.\ Alkofer at Eighth International Symposium on
Meson-Nucleon Physics and the structure of the nucleon (MENU99),
Zuoz, Engadine, Switzerland August 15-21, 1999. }} 
\author{R.~Alkofer, S.~Ahlig, C.~Fischer, and M.~Oettel \\
{\em Institute for Theoretical Physics, University of T\"ubingen,\\
Auf der Morgenstelle 14, D-72076 T{\"u}bingen}}

\maketitle

\begin{abstract}
\setlength{\baselineskip}{2.6ex}
Baryons are modeled as bound states of scalar or axialvector diquarks and
a constituent quark which interact through quark exchange.
This description results as an approximation to the relativistic
Faddeev equation for three quarks. The corresponding effective Bethe-Salpeter
equation is solved, and fully four-dimensional wave functions for both octet 
and decuplet baryons are obtained. Confinement of quarks and diquarks is
incorporated by modifying their propagators. Results are given for kaon 
photoproduction which is one of the applications currently
under investigation in this model.
\end{abstract}

\setlength{\baselineskip}{2.6ex}

\section*{Motivation}

In this talk it will be argued that in the study of baryonic structure
kaon photoproduction is an especially interesting hadronic reaction.
Photo-- and electroproduction of pseudoscalar mesons have been studied
intensively in isobar resonance models ~\cite{Hab99,Tia99} and coupled
channel calculations of hadronic interactions ~\cite{Kai99}. These models are
capable of describing the wealth of measured data reasonably well. 
However, by construction they do not allow for an interpretation of the
production process in terms of baryon substructure. The aim of the
investigation reported here is to clarify whether the notion of diquarks as
effective constituents of baryons in meson production close to threshold
is helpful in the understanding of subnucleonic physics.

Approaches to the substructure of baryons include nonrelativistic quark models,
various sorts of bag models and different types of solitons ~\cite{Bhad88}.
Most of these models are designed to work in the low energy region and
generally do not match the  calculations within perturbative QCD. During this
conference impressive reports on the great experimental progress in medium
energy physics have been given. This underlines the high demand for models
describing baryon physics in this region. Therefore
we propose a fully covariant description of baryon structure in the framework
of a diquark--quark--model of baryons.

Our motivation to choose such an approach is based on two sources. On the one
hand, when starting with the fully relativistic Faddeev equation for bound
states of three quarks, diquarks appear as effective degrees of freedom. These
diquarks stand for (potentially weakly)  
correlated quark--quark pairs inside baryons. 
On the other hand, diquarks as constituents of
baryons are naturally obtained when one starts with an NJL--type  model of
colour octet flavour singlet quark currents ~\cite{alk95}. Although in the limit
$N_c \rightarrow \infty$ baryons emerge as solitons of meson fields
~\cite{alk96}, it can be shown for the case of three colours that both effects,
binding through quark exchange in the diquark-quark picture and through mesonic
effects, contribute equally ~\cite{zuc96}.

\section*{The four-dimensional Bethe-Salpeter equation: masses and 
wave functions}

As stated above, in a first step we want to reduce the complexity 
of the full three--body problem of the relativistic Faddeev equation
for baryons.  
This can be achieved by approximating the two--quark
irrreducible $T$--matrix by separable contributions that can be viewed as
loosely bound diquarks. The three--body problem then 
becomes an effective two--body
one, in which bound states appear as the solution of a homogeneous
Bethe--Salpeter equation. The attractive interaction between quark and
diquark is hereby provided by quark exchange. This interaction is but
the minimal correlation needed to reconstitute the Pauli principle.
Due to antisymmetry in the color indices and the related
symmetrization of all other quantum numbers the Pauli principle
leads to an attractive interaction in contrast to ''Pauli
repulsion'' known in conventional few-fermion systems.
All unknown and probably very complicated gluonic interactions between two
quarks are effectively treated via the parameterization of the diquark propagator
and the diquark--quark--quark vertex function.
In Ref.\ ~\cite{oett98} we have formalized this procedure by an effective Lagrangian
containing constituent quark, scalar diquark and axialvector diquark fields.
This leads to a coupled set of Bethe--Salpeter equations for octet and
decuplet baryons.

\begin{table}[t]
\caption{Components of the octet baryon wave function with their
respective spin and orbital angular momentum.
$(\gamma_5 C)$ corresponds to
scalar and $(\gamma^\mu C),\,\mu=1 \dots 4,$ to axialvector
diquark correlations. Note that the partial waves in the first row
possess a non-relativistic limit. See ~\cite{oett98} for
further details.}
\begin{tabular}{p{0.1cm}p{1.8cm}cccc}\hline
 &&&&& \\
\multicolumn{2}{l}{\parbox{1.9cm}{ { \mbox{``non-relat.''} \mbox{partial waves}} }  } &
  $\pmatrix{ \st \chi \cr \st 0 } \st{(\gamma_5 C)}$    &
  $\st\hat P^4{\pmatrix{ \st 0\cr \st \chi}} (\gamma^4 C)$    &
  ${\pmatrix{\st i\sigma^i\chi \cr \st 0}} \st (\gamma^i C)$    &
  ${\pmatrix{\st i\left(\hat p^i(\vec{\sigma}\hat{\vec{p}})-\frac{\sigma^i}{3}\right) \chi\cr \st 0}} \st (\gamma^i C)$ \\  
 &  {spin} & {1/2} & {1/2} & {1/2} & {3/2} \\
 &  {orb.ang.mom.} & { $s$} & {$s$} & {$s$} & {$d$} \\
 &                  &        &       &       &      \\
\multicolumn{2}{l}{\parbox{1.9cm}{ { \mbox{``relat.''} \mbox{partial waves} }  }} &
  $\pmatrix{ \st 0 \cr \st \vec \sigma \vec p \chi } \st (\gamma_5 C)$    &
  $\st \hat P^4{\pmatrix{ \st (\vec{\sigma}\vec{p})\chi\cr \st 0}}(\gamma^4 C)$    &
  ${\pmatrix{\st 0\cr \st i\sigma^i(\vec{\sigma}\vec{p})\chi}}\st (\gamma^i C)$    &
  ${\pmatrix{\st 0\cr \st i\left(p^i-\frac{\sigma^i(\vec{\sigma}\vec{p})}{3}\right)\chi}} \st (\gamma^i C)$ \\   
 &  {spin} & \cl{1/2} & \cl{1/2} & \cl{1/2} & {3/2} \\
 & {orb.ang.mom.} & \cl{ $p$} & \cl{$p$} & \cl{$p$} & {$p$}\\ \hline
\end{tabular}
\label{wave}
\end{table}

\begin{table}[b]
\caption{Octet and decuplet masses obtained with two different parameter sets.
Set I represents a calculation with weakly confining propagators, Set II
with strongly confining propagators, see ~\cite{oett98}.
All masses are given in GeV.}
\begin{tabularx}{\linewidth}{YYYYYYYYY} \hline
 & $m_u$ & $m_s$ & $M_\Lambda$ & $M_\Sigma$ & $M_\Xi$ & $M_{\Sigma^*}$ & $M_{\Xi^*}$ & $M_\Omega$ \\ \hline
% &&&&&&&& \\
Set I & 0.5 & 0.65 & 1.123 & 1.134 & 1.307 & 1.373 & 1.545 & 1.692 \\
Set II& 0.5 & 0.63 & 1.133 & 1.140 & 1.319 & 1.380 & 1.516 & 1.665 \\
Exp.  &     &      & 1.116 & 1.193 & 1.315 & 1.384 & 1.530 & 1.672 \\ 
% &&&&&&&& \\ 
\hline
\end{tabularx}
\label{masses}
\end{table}

We avoid unphysical thresholds by an effective parameterization of
confinement in the quark and diquark propagators.
We solve the four--dimensional equations
in ladder approximation and obtain wave functions for the octet and
decuplet baryons ~\cite{oett98}.
The Lorentz invariance
of our model has been checked explicitly by choosing different frames.

The implementation of the appropriate Dirac  and Lorentz
representations
of the quark and diquark parts of the wave functions leads to a unique
decomposition in the rest frame of the baryon. Besides the well known $s$-wave and
$d$-wave components of non-relativistic formulations of the baryon octet
we additionally obtain
non-negligible $p$-wave contributions which demonstrates again the need for
covariantly constructed models. Table \ref{wave} summarizes the
structure of the octet wave function. Each of the eight components is to be
multiplied with a scalar function which is given in terms of an expansion
in hyperspherical harmonics and is computed numerically.

In order to obtain the mass spectra for the
octet and decuplet baryons we explicitly break SU(3) flavour symmetry by
a higher strange quark constituent mass.
Using the nucleon and the delta mass as input our calculated mass spectra 
~\cite {oett98} are in good
agreement with the experimental ones, see Table \ref{masses}.
The wave functions for baryons with distinct
strangeness content but same spin differ mostly due to
flavour Clebsch-Gordan coefficients, the respective invariant functions being
very similar.
Due to its special
role among the other baryons, we investigated the
$\Lambda$ hyperon in more detail and discussed its vertex amplitudes.
In our approach, the $\Lambda$ acquires a small flavour singlet
admixture which is absent in $SU(6)$ symmetric
non-relativistic quark models.

\section*{Form Factors}

A significant test of our model is the calculation of
various form factors ~\cite{hell97,oet99,ahlig99}\footnote{For a similar
calculation of the electromagnetic nucleon form factors within a slightly
different diquark model see Ref.\ ~\cite{Ben99}}.  The most important ingredient
are the fully four-dimensonal wave functions described above. It turns out that
already the electromagnetic form factors of the nucleon provide severe
restrictions for the parameters of  the model.

For the pion--nucleon form factor at the soft point, $g_{\pi NN}$, we find 
good agreement with experiment. For spacelike momenta this form factor falls
like a monopole with a large cutoff similar to the behaviour in
One-Boson-Exchange (OBE) models ~\cite{ahlig99}.  Compared with a calculation
including only scalar diquarks ~\cite{hell97} we find a lower value for the
pion--nucleon coupling at the soft point. Serving as a central ingredient for
strangeness production processes the kaon--nucleon--lambda form factor
$g_{KN\Lambda}$ is  a quantity of special interest. Due to flavour algebra the
isospin configuration of the $\Lambda$ singles out the scalar diquark as the
only diquark contributing to nucleon--lambda transitions.  Since a pseudoscalar
kaon does
not couple to the scalar diquark we find ourselves in a comfortable position
to handle such transitions. We find that the absolute value of the
kaon--nucleon--lambda form factor is always smaller than the pion--nucleon form
factor ~ \cite{ahlig99}. This can be understood from the facts that the
axialvector diquark does not contribute and that the kaon decay constant is
larger than the pion decay constant. 

\begin{figure}[t]
\begin{center}
\epsfig{file=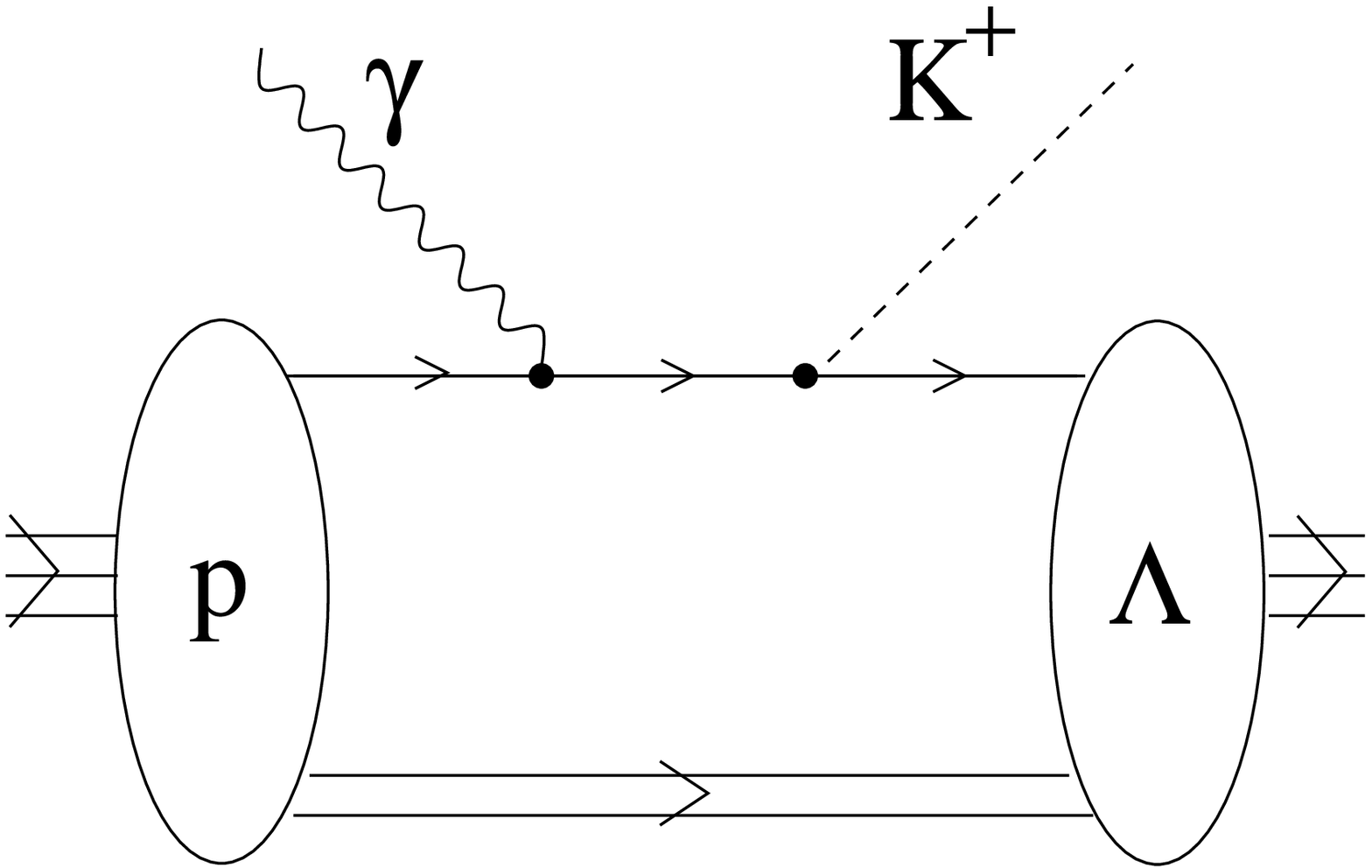,width=4.3cm} \hspace{2.6cm}
\epsfig{file=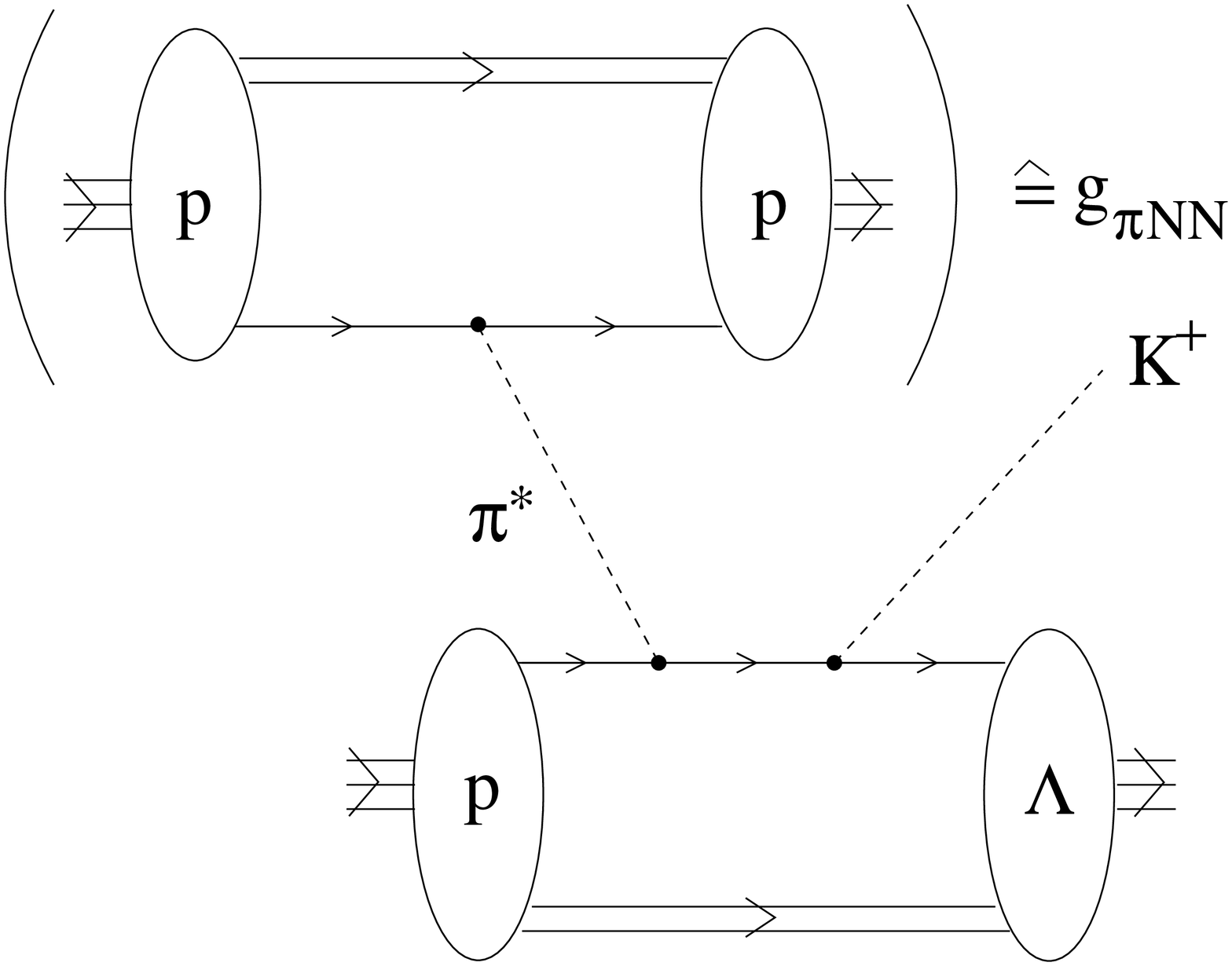,width=4.3cm}
\end{center}
\vspace{-5mm}
\caption{Typical diagrams to be computed in the diquark
spectator picture. Left: kaon photoproduction. Right:
strangeness production in proton--proton collisions.}
\label{pics}
\end{figure}

\section*{Kaon Photoproduction}

The SAPHIR collaboration has measured the cross section and the asymmetries for
the process $p\gamma \rightarrow K\Lambda$ with high precision. These
measurements continue to prompt corresponding calculations within various
models, among these are the isobaric model ~\cite{Hab99,Tia99}, the coupled
channel approach ~\cite{Kai99} and models which describe those processes in
terms of the dynamics of the baryon substructure ~\cite{Kro96}.

The reaction $p\gamma \rightarrow K\Lambda$ lends itself to a description within
the diquark--quark picture. This is for two reasons which greatly simplify the
description: first, the scalar diquark is the sole overlap between the
wave function of the proton and the lambda, which implies that axialvector
diquarks cannot participate in the reaction, and second, the kaon does
not couple to the diquark. This leads to the conclusion that the diagram
shown in the left half of Fig.\ \ref{pics} and the corresponding
crossed diagram are the dominant contributions to kaon photoproduction.

\begin{figure}[t]
\begin{center}
\epsfig{file=saphir.eps,width=0.495\linewidth}
\epsfig{file=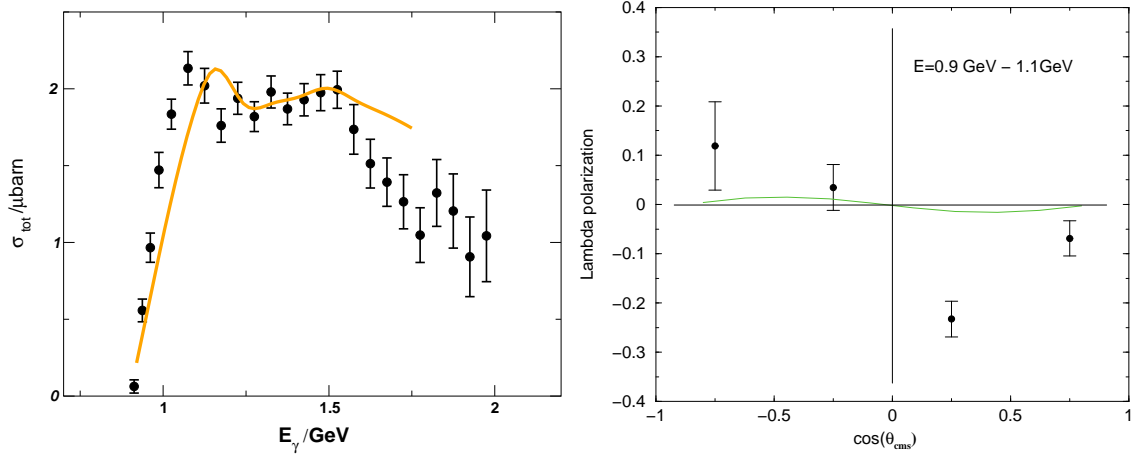,width=0.495\linewidth}
\end{center}
\caption{Total cross section (left panel) and Lambda polarization (right panel)
for the reaction $p\gamma \rightarrow K\Lambda$.
The data are taken from ~\cite{SAPHIR} and are shown together with our results.}
\label{total}
\end{figure}

The total cross section for $p\gamma \rightarrow K\Lambda$ is shown in the left
panel of Fig.\ \ref{total}. For $E < 1.5$ GeV the data are reproduced quite
nicely, whereas our results do not fall off fast enough for higher energies.
We found that the total cross section depends very little on the details of
the baryon wave functions, however, we observed a rather pronounced dependence on
the way confinement is parameterized into the propagators.
A closer analysis reveals that production processes like
$p\gamma \rightarrow K\Lambda$
probe the propagators of the constituents in the timelike region, i.e.
for timelike momenta of the constituents. This is a
highly welcome feature which may be used in conjunction with Dyson--Schwinger
studies. This approach is tied to the dynamics of QCD and thus gives access
to the nonperturbative quark propagator, which enters as a key ingredient
in the description of baryons as bound states of quarks and diquarks.
Since the Dyson-Schwinger approach gives the propagators for
spacelike momenta only, one should take on board the idea, that
production processes like $p\gamma \rightarrow K\Lambda$ firmly constrain
the extrapolation to timelike momenta.

The lambda polarization is shown in the right panel of Fig.\ \ref{total} for the
energy range $0.9-1.1$ GeV. Our results fall short of the experimental values,
however, the change in sign is reproduced. Comparing with, e.g. 
~\cite{Hab99,Tia99,Kai99}
we conclude that apparently none of the current models is able to reproduce
the asymmetries to reasonable accuracy.

\vspace{1cm}
{\bf Acknowledgement:}
R.A.\ thanks the organizers for making this conference so pleasant and 
stimulating.  The authors  also want
to express their gratitude to Hugo Reinhardt and Herbert Weigel
for their support of this project.

\bibliographystyle{unsrt}

\end{document}